\documentclass[12pt]{article}

\usepackage{geometry}
\usepackage{amsmath,amssymb,amsfonts}
\usepackage{mathtools}
\usepackage{bbold}
\usepackage{hyperref}

\usepackage[dvipsnames]{xcolor}

\usepackage{algorithm}
\usepackage{caption}
\usepackage{romannum}
\usepackage{marginnote}
\usepackage{mparhack}
\usepackage{upgreek}

\begin{document}

\begin{titlepage}
   \begin{center}

       \section*{Optimized encoding point distributions for efficient single-point imaging}
       \vspace{0.5cm}
       \text{Fabian Bschorr$^{1,2}$, Pia Gebhard$^{1,2}$, Tobias Speidel$^{1}$, Volker Rasche$^{1,3}$}\\
       \vspace{0.2cm}
       $^{1}$ Department of Internal Medicine II, Ulm University Medical Center, Albert-Einstein-Allee 23, 89081 Ulm, Baden-Württemberg, Germany \\
       $^{2}$ Author Pia Gebhard and Fabian Bschorr contributed equally to this work\\ 
       $^{3}$ Core Facility Small Animal MRI (CF-SANI), Albert-Einstein-Allee 11, 89081 Ulm, Baden-Württemberg, Germany \\                
   \end{center}
   
   \subsection*{Abstract}
   
   \textbf{Purpose}: Quasi-random Sobol–based sampling schemes exhibit deterministic structural artifacts when aggressively undersampled, particularly at low encoding densities required for accelerated 2D SPI/CSI. To address these limitations, two advanced undersampling strategies are investigated to mitigate deterministic behavior improving image quality for time-constrained applications such as hyperpolarized MRI.\\ 
   \textbf{Methods}: An optimized Sobol sequence-derived point distribution with Heaviside-type density gradient center oversampling served as the initial sampling pattern. Undersampling was performed using two point-reduction algorithms: radius-adaptive stochastic undersampling (RAST), which applies a geometric, radius-dependent minimum-distance criterion, and Bayesian Information Gain Optimization (BINGO), that removes points based on their information gain to the reconstructed image. 
   Phantom experiments were conducted on a 3 T clinical MRI system using up to 16-fold undersampling.\\ 
   Image quality was quantified using a performance score derived from RMSE, SSIM, and HFEN. 
   \textbf{Results}: Both RAST and BINGO outperformed deterministic undersampling across all metrics. RAST achieved highest and most robust performance, with improvements up to 238\% in the averaged metric score, while BINGO yielded improvements of 133\% across matrix resolutions. \\
   \textbf{Conclusion}: The proposed strategies effectively reduce the number of encoding points in low-discrepancy 2D SPI point distributions while maintaining image quality under strong acceleration. RAST provides superior 
   metric performance, whereas BINGO offers broad applicability, including suitability for non-linear encoding fields. These approaches support rapid acquisition workflows required for real-time and hyperpolarized applications.
\end{titlepage}

\section{Introduction}\label{sec1}

Magnetic resonance imaging (MRI) is a cornerstone of modern diagnostic imaging, providing excellent soft tissue contrast and functional information without ionizing radiation. However, the inherent long acquisition times of conventional MRI limit its use in time-critical applications such as hyperpolarized, real-time, or metabolic imaging \cite{WODTKE20258}. In such scenarios, rapid image acquisition is crucial for capturing transient signals or dynamic processes while maintaining clinically acceptable image quality.

The increasing interest in multinuclear imaging, especially in the context of $^{13}$C for hyperpolarized imaging driven by recent developments in easy-to-use hyperpolarization devices \cite{nagel2023PHIP, gierse23fumarate}, raises interest in frequency-resolved image acquisitions with 2D chemical shift imaging (CSI) being an important workhorse. Due to limited life time and/or low sensitivities of the compounds, matrix sizes used in research barely exceed $64\times 64$ \cite{CSSI2D_paper1, CSI2D_paper2, CSI2D_paper3}.

The coherent aliasing properties of Cartesian sampling causes distinct and spatially correlated artifacts in case of undersampling. Even though classical, non-Cartesian encoding, such as radial or spiral, shows more favorable undersampling properties, low-discrepancy point distributions \cite{CSLustig}, which distribute sampling points quasi-randomly across k-space have been shown to be advantageous. They offer incoherent, noise-like aliasing that can be effectively suppressed using advanced reconstruction methods such as compressed sensing or parallel imaging \cite{Dwork, Foucart}.

Recent developments in low-discrepancy sampling design have further proven the potential for enabling high-undersampling at decent image quality. Such point distributions balance coverage uniformity and sampling randomness, causing predominantly non-structured (noise-like) aliasing artifacts and are thus favorable for reconstruction from highly undersampled data \cite{TobiPaper, Rasche}. 

Despite the potential to accelerate image acquisition, point clouds generated by, e.g., Sobol sequences become deterministic at low point densities \cite{RENARDY2021108593, Paulin10.1145/3478513.3480482}. Thus, optimization of 2D CSI with small matrix sizes becomes challenging, demanding optimization schemes for reducing the number of sample points in an initially fully sampled, low-discrepancy point distribution.

Importantly, while conventional image quality metrics, such as signal-to-noise ratio (SNR) or root-mean-square error (RMSE), provide quantitative benchmarks, they may not fully reflect clinical usability \cite{Ravishankar5617283}. For time-critical MRI applications, diagnostic value and interpretability, rather than traditional measures of image fidelity, are the ultimate criteria for applicability and diagnostic image quality. 


In this work, we investigate acceleration strategies for MRI in time-critical settings using non-Cartesian, low-discrepancy sampling schemes, especially for small matrix sizes, by employing two different algorithms to reduce the number of encoding steps starting from an initial non-Cartesian low-discrepancy point distribution. We assess their potential to generate useful images at high acceleration factors and discuss implications for real-time and hyperpolarization imaging workflows.


\section{Theory} \label{sec:Theory}
In the following, two algorithms are presented for further reduction of encoding steps based on an initial Sobol distribution.

\subsection{Initial point distribution} \label{sec:InitTraj}
In previous works \cite{ISMRMPia}, we introduced an optimization strategy for two-dimensional single-point imaging (SPI) based on  a Sobol base-2 sequence \cite{Sobol}. Building on this foundation, we incorporated center oversampling (COS) and performed PSF-based optimization to identify density functions that minimize the side-lobe-to-peak ratio (SPR) \cite{CSLustig, Hao1415445}. This process ultimately favored a Heaviside-type density distribution, which substantially reduces the SPR. Together, these steps provide an optimized point set that serves as an ideal starting point for this work.

\subsection{Geometric Optimization}
Straightforward undersampling of the Sobol point set to accelerate the acquisition, i.e. by consideration of every $n$-th point only, fails at low point densities. Since the algorithm for the generation of the Sobol-sequence is inherently deterministic \cite{JoeKuo, Sobol2}, aggressive undersampling amplifies its underlying structural regularities, destroying the uniformity and low-discrepancy properties. As the point set becomes increasingly sparse, coherent and directional aliasing dominates, rather than the desired noise-like artifact behavior. Consequently, deterministic undersampling is not suitable for accelerating acquisitions in regimes where only a small number of encoding steps are available. Instead, an adaptive undersampling strategy is introduced, which is in the following referred to as radius-adaptive stochastic thinning (RAST). The two-dimensional point set $X$ is geometrically undersampled by evaluating the individual point radius $r_i$ of each point. The points are processed in random order using a point-picking strategy $\pi(i)$ and are accepted only if they satisfy a radius-dependent minimum-distance criterion relative to the subset of points already selected. To ensure that all points in the set are evaluated, a positive, radius-dependent spacing $r_{\text{min}}$ is imposed at the start of the procedure:
\begin{equation}
	\tilde{X} = \left\{ x_i \in X : ||x_{\pi(i)} - x_{\pi(j)}|| > r_{\text{min} }(r_i),  \forall i\neq j \in |X|\right\}.
\end{equation}
This approach preserves the original density gradient while introducing stochasticity that promotes quasi-uniform, blue-noise-like behavior towards the periphery \cite{Heck10.1145/2487228.2487233, Ulichney3288}. As a result, geometric undersampling avoids the visible coherent and structural regularities of deterministic undersampling, reduces structural artifacts, and improves overall spatial uniformity.

\subsection{Bayesian Information Gain Optimization}
As an alternative to the geometric approach, we also include an algorithm not assuming any specific encoding fields, e.g. linear gradients. To evaluate the encoding capabilities of different magnetic field configurations, which can, of course, include both classical point distributions and non-linear encoding schemes, different quantitative metrics have already been utilized \cite{trajOptNon, seeger2010optimization}. Inspired by the works on optimizing point distributions, i.e. evaluating potential candidates for appending to the point distribution based on their performance after compressed sensing \cite{lustig2008compressed}, we propose an optimization algorithm that only relies on the encoding matrix. 

\paragraph{Signal Equation} The signal $\vec{s}$ of and object function $\vec{\rho}$ for an arbitrary MRI sequence is given by \cite{fessler2010model}
\begin{equation}
	\vec{s} = E \vec{\rho} + \vec{\varepsilon},
\end{equation}
where $E$ is the encoding matrix with Gaussian distributed noise $\varepsilon \in \mathcal{N}(0, \sigma^2)$ with variance $\sigma^2$.

\paragraph{Reconstruction} The probability density for measuring $\vec{s}$, having the object $\vec{\rho}$ is also Gaussian distributed
\begin{equation}
	p(\vec{s}\,|\,\vec{\rho}) = \mathcal{N}(\vec{s}; E\vec{\rho}, \sigma^2 \mathbb{1}).
\end{equation}
With this and assuming a Gaussian prior, the posterior distribution function can be calculated with Bayes' theorem according to

\begin{align}
	p(\vec{\rho}\,|\,\vec{s}) &= \frac{p(\vec{s}\,|\,\vec{\rho}) p(\vec{\rho})}{p(\vec{s})}\\
    &\propto \mathcal{N}(\vec{s}; E \vec{\rho}, \sigma^2 \mathbb{1}) \mathcal{N}(\vec{\rho}; \vec{0},P_0) \nonumber\\
	&\propto \exp\left(-\frac{1}{2} (\vec{\rho}-A^{-1}\vec{b})^H A (\vec{\rho}-A^{-1}\vec{b}) \right)\\ 
    &= \mathcal{N}(\vec{\rho}; A^{-1}\vec{b}, A^{-1}),
\end{align}
where
\begin{align}
	A = \frac{E^H E}{\sigma^2} + \frac{1}{P_0^2} \,, \, \vec{b} = \frac{1}{\sigma^2} E^H \vec{\rho}.
\end{align}
In these equations, $p(\vec{\rho})$ serves as prior information about the image/magnetization distribution, e.g. being sparse. However, for this work, we will not assume any prior information later on by letting $P_0 \rightarrow \infty$.

Later, the optimization task involves the decision, which encoding step is least important to a certain point distribution. Thus, motivated by Seeger et al.\cite{seeger2010optimization}, the design score here will be based on the information gain calculated by the difference in the differential entropies of the posterior distribution with and without an additional encoding step. For a Gaussian, the differential entropy $H$ reduces to
\begin{equation}
	H(p(\vec{\rho}\,|\,\vec{s})) = \frac{1}{2} \ln{|2\pi e A^{-1}|}
\end{equation}
where $|.|$ is the determinant of the matrix.\\
Considering one additional encoding step of the point distribution by appending the respective rows to the encoding matrix $E$ yields $\tilde{E}$, such that the signal equation is given by
\begin{equation}
	\tilde{\vec{s}} = \tilde{E} \vec{\rho} + \vec{\varepsilon}.
\end{equation}
The information gain $\mathcal{I}$ can then be evaluated by comparing $p(\vec{\rho}|\vec{s})$ with $p(\vec{\rho}|\tilde{\vec{s}})$:
\begin{align}
	\mathcal{I} &= H(p(\vec{\rho}\,|\,\vec{s}))- H(p(\vec{\rho}\,|\,\tilde{\vec{s}})) = \nonumber \\ &=\frac{1}{2} \ln{|2\pi e A^{-1}|} - \frac{1}{2} \ln{|2\pi e \tilde{A}^{-1}|} = \nonumber \\
	&= \frac{1}{2} \ln{\frac{|2\pi e A^{-1}|}{|2\pi e \tilde{A}^{-1}|}} = \frac{1}{2} \ln{\frac{|\tilde{A}|}{|A|}}.
\end{align}
In principle, we are now ready to proceed with the optimization, however, the numerical implementation of this algorithm involves additional work for a reliable and fast implementation. Since $E$ and $\tilde{E}$ are directly connected to each other due to removal of only one encoding step, the equation can be simplified. Let $r_i$ be the i-th row of $\tilde{E}$, then, we can write
\begin{equation}
	\tilde{E}^H \tilde{E} = \sum_{j=1}^{T \zeta} r_j r_j^H, E^H E = \sum_{j=1, i\neq j}^{T \zeta} r_j r_j^H,
\end{equation}
with $T$ being the number of time steps and $\zeta$ the number of encodings used such that
\begin{equation}
	\tilde{A} = \tilde{E}^H\tilde{E} = A + r_i r_i^H.
\end{equation}
With this equation and the help of the matrix determinant lemma \cite{matrixdetLemma}, a more efficient way of writing the information gain is provided by
\begin{equation}
	\mathcal{I} = - \frac{1}{2} \ln{\left( 1- r_i^H A^{-1} r_i \right)}.
\end{equation}
The presented algorithm, which is based on Bayes' information gain optimization, is hereafter denoted as BINGO. It only relies on the encoding matrix, thus, being not limited to classical, linear encoding schemes.

\section{Methods}\label{sec:meth}
\begin{figure}[htbp]
	{%
		\centering
		\includegraphics[width=0.3\linewidth]{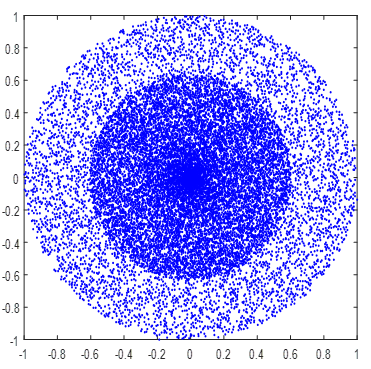}
		\captionof{figure}{%
			Optimized initial point distribution based on a Sobol base-2 sequence and center oversampling with Heaviside density function with width of 62\% radius on the unit circle.}
		\label{fig:InitTraj}
	}
\end{figure}

\begin{figure}[htbp]
	{%
		\centering
		\includegraphics[width=0.7\linewidth]{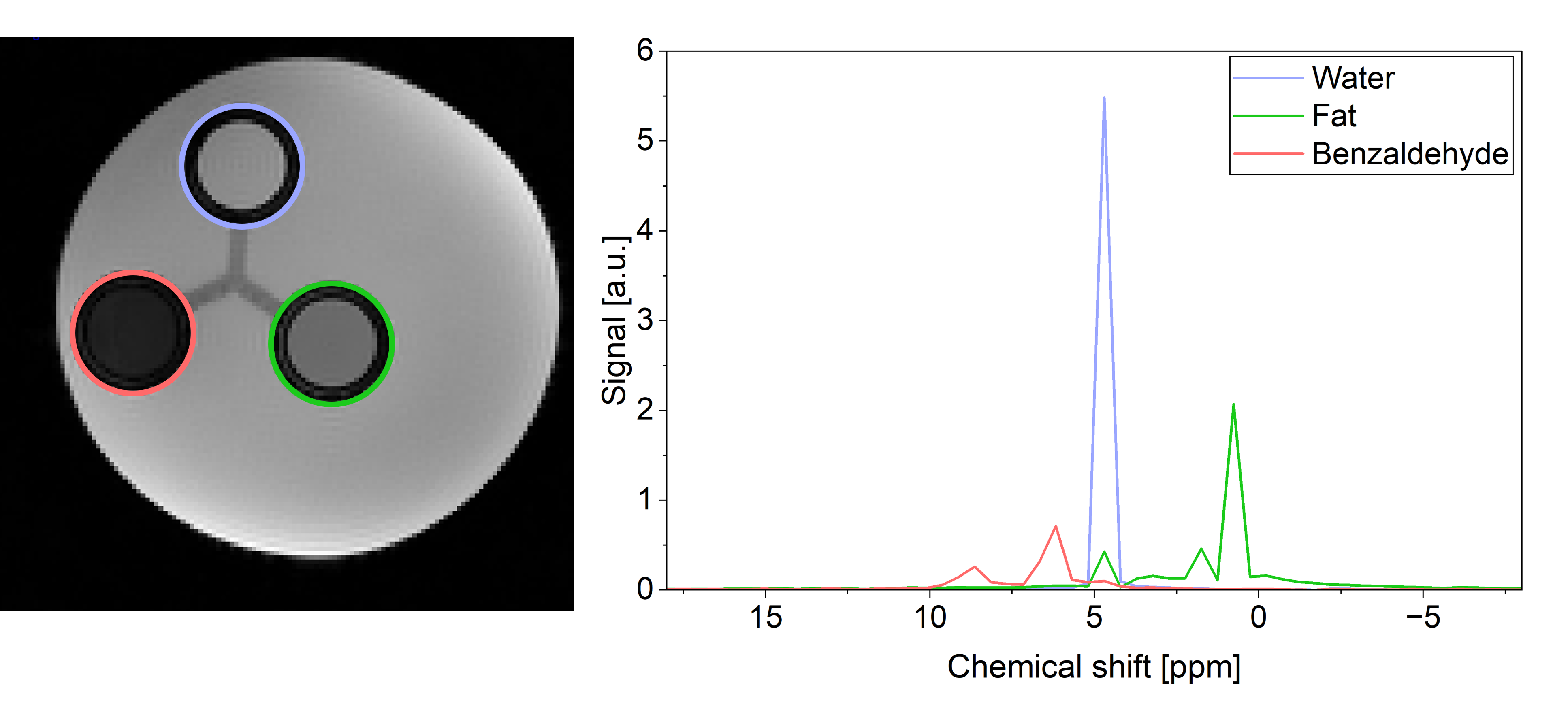}
		\captionof{figure}{%
			Custom-built phantom consisting of three compartments filled with isotonic saline solution (water, blue), fat (green), and benzaldehyde (red). The corresponding chemical shift imaging spectrum acquired from the phantom is shown on the right.}
		\label{fig:phantom_spectrum}
	}
\end{figure}

The presented 2D SPI optimization strategies were experimentally validated using a stationary, custom-built phantom containing three identical in-house–assembled glass vials, filled with 0.9\% NaCl solution containing 0.5 mmol/ml Dotarem\textsuperscript{\textregistered} contrast agent, sunflower oil, and benzaldehyde, with remaining air bubbles carefully removed. The vials are embedded in an 1\,l container, filled with the same isotonic phantom solution to ensure a homogeneous background. A CSI-acquired spectrum of all metabolites, as well as the acquired 2D SPI image of the phantom, is shown in fig. \ref{fig:phantom_spectrum}.
The initial point distribution was generated according to the procedure described in previous works \cite{ISMRMPia} and in sec. \ref{sec:InitTraj}, and is shown in fig.\,\ref{fig:InitTraj}\!. 
2D SPI/CSI experiments were performed on a 3.0\,T whole-body MRI system (Ingenia 3.0 T CX, Philips, Best, The Netherlands) using a single-loop DStream Flex-M coil (Philips Healthcare). For large matrix evaluation, data were acquired with a $128 \times 128$ imaging matrix over a $( 128 \times 128 \times 5)$ mm$^3$ field of view. 
The flip angle was set to 15$^{\circ}$ based on the Ernst angle of the phantom components, with $T_R = 23.98$ ms, acquisition time $T_{\mathrm{acq}} = 15.90$ ms, and scan duration of $\leq$ 6:30 min. Spectral separation was 62.88 Hz with a sample bandwidth of 8.04 kHz.

For comparison of deterministic undersampling with the two proposed algorithms 8-, 12- and 16-fold undersampling was evaluated for an imaging matrix of $128 \times 128$. 
For BINGO, the whole encoding matrix has to be set up leading to large matrix sizes, since $E \in \mathbb{C}^{n \cdot u \cdot c \times D^2}$ with $n$ time-steps, $u$ encoding steps, $c$ receiver coils, and $D^2$ spatial evaluation points. For reasonable evaluation times, $D^2$ was fixed to $32^2$ instead of $128^2$. 
For each point in the point cloud, the information gain of every encoding step was evaluated and the point with the smallest information gain removed from the point distribution.

To further evaluate the performance of both algorithms for smaller matrix sizes, the same experiment was conducted with a $32 \times 32$ imaging matrix and $D^2=32^2$ leading to $1024$ encoding steps and scan duration $\leq 25$\,s. Similarly, deterministic undersampling was compared with the proposed methods using 2-, 4- and 8-fold undersampling.
Reconstruction involved a non-uniform FFT in the BART reconstruction framework \cite{uecker2015berkeley} for full resolution imaging matrix of size $128 \times 128$ and low-resolution $32\times32$ matrix. In addition, compressed sensing (CS) reconstruction with total variation regularization was employed to mitigate the noise-like aliasing artifacts inherent to the undersampled data.

To evaluate the performance of the introduced undersampling strategies, three metrics were employed: normalized root mean square error (RMSE), structure similarity index measure (SSIM), and normalized high-frequency error norm (HFEN) \cite{Ravishankar5617283, Mittal6272356}. For each undersampling pattern and each undersampling factor, the corresponding image metric was computed. To quantify the overall performance of an undersampling pattern, a performance score was defined as the mean of the normalized metrics. This score was calculated for both the high-resolution $128 \times 128$ images and the $32 \times 32$ matrix images, followed by an average score derived from both resolutions.

\section{Results}

\begin{figure*}[htbp]
	{%
		\centering
		\includegraphics[width=1\linewidth]{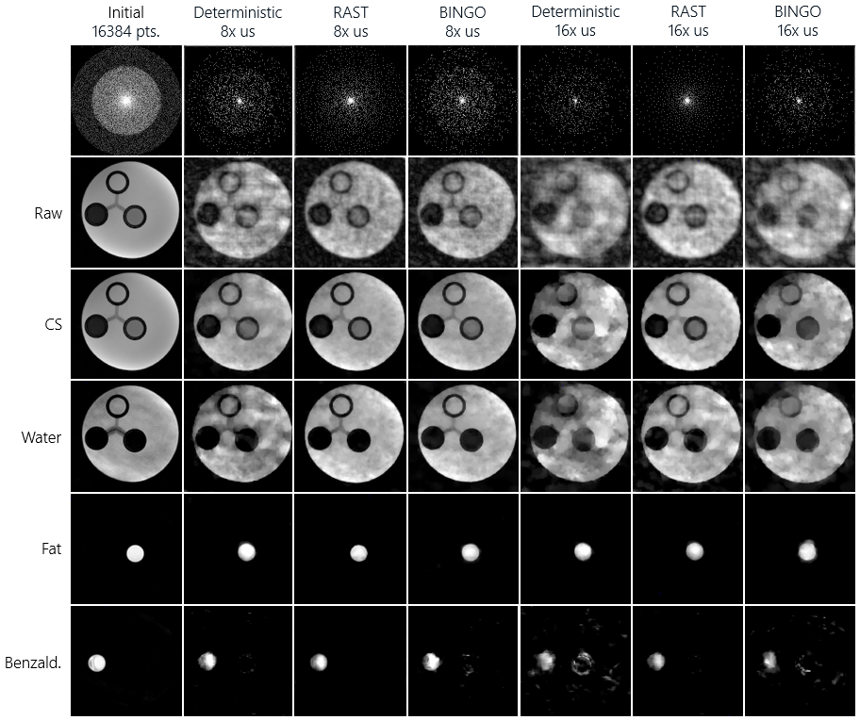}
		\captionof{figure}{%
			Comparison of undersampling strategies applied to the initial point distribution (top row): deterministic, RAST, and BINGO, for reconstructed matrix size of $128 \times 128$. Raw reconstructions and corresponding compressed sensing–reconstructed images (CS) are displayed for 8-fold (2048 points) and 16-fold (1024 points) undersampling. The bottom three rows show the chemical shift separated images for the phantom compartments, water, fat, and benzaldehyde, respectively (after CS).}
		\label{fig:US128}
	}
\end{figure*}

\begin{figure*}[htbp]
	{%
		\centering
		\includegraphics[width=1\linewidth]{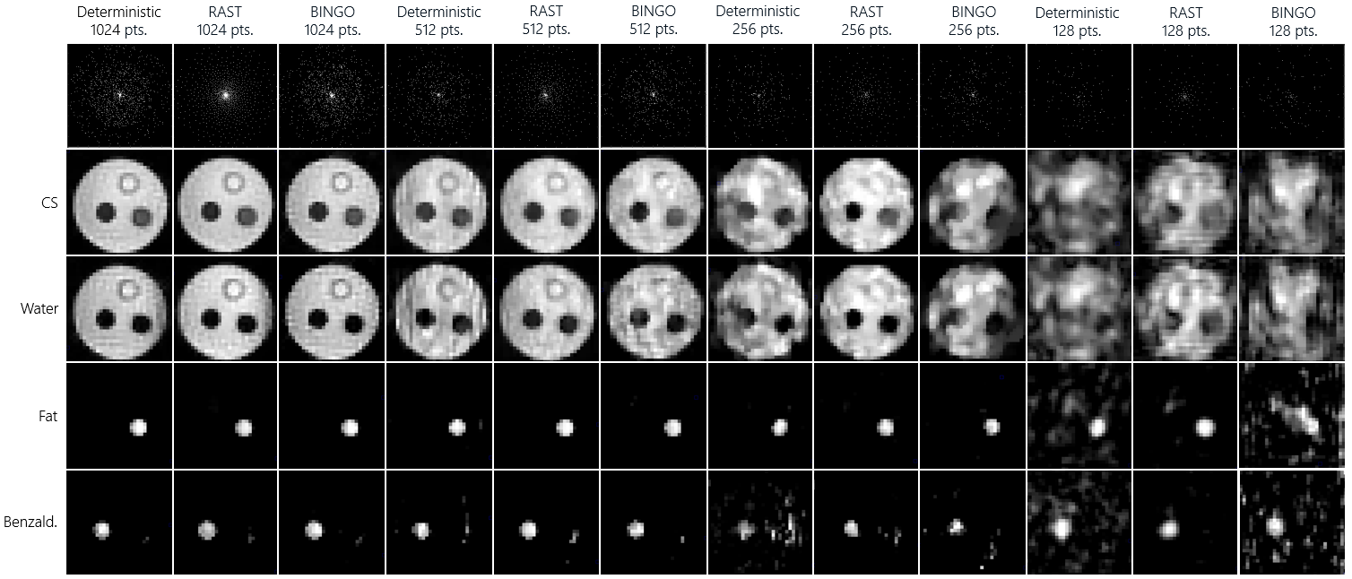}
		\captionof{figure}{%
		      Sampling patterns for a $32\times32$ (1024 points) imaging matrix, 2-fold (512 points), 4-fold (256 points) and 8-fold (128 points) undersampling for deterministic, RAST, and BINGO undersampling with corresponding compressed sensing–reconstructed images (CS). In addition, chemical shift separated images (CSI) of phantom compartments, water, fat, and benzaldehyde are shown, respectively (after CS).}
		\label{fig:US32}
	}
\end{figure*}

The resulting optimized point distributions for $128^2 = 16,384$ points are shown in the top row of fig. \ref{fig:US128}, exemplarily for 8-fold and 16-fold undersampling, reducing the initial point distribution to 2048 and 1024 points, respectively. The corresponding raw and CS reconstructed images as well as the separated metabolite images for an imaging matrix size of $128\times128$ are displayed alongside the respective point distributions and compared to the fully sampled reference shown on the left. Figure \ref{fig:US32} presents the CS-reconstructed images, separated metabolite images, and corresponding point distributions for a reduced imaging matrix size of $32 \times 32$, including 1024 points as well as 
2-fold, 4-fold, and 8-fold undersampling, approaching point counts as low as 128.
Deterministic undersampling results in coherent, patchy artifacts which cannot be sufficiently mitigated by CS, particularly at low point densities. BINGO exhibits similar limitations at very sparse point densities and retains structured artifacts, most notably for 128 and 256 points. In contrast, RAST consistently yields predominantly incoherent undersampling artifacts, which are more effectively suppressed by CS reconstruction and result in improved image quality. The separated water images show similar artifact behavior as the CS-reconstructed images whereas the spectral separation for fat works well for the shown undersampling factors. Benzaldehyde, despite having substantially lower signal intensity compared to water, can also be separated from the other phantom components, whereas for low point densities, structured artifacts and increased bleeding arise.

\begin{figure}[htbp]
	{%
		\centering
		\includegraphics[width=0.6\linewidth]{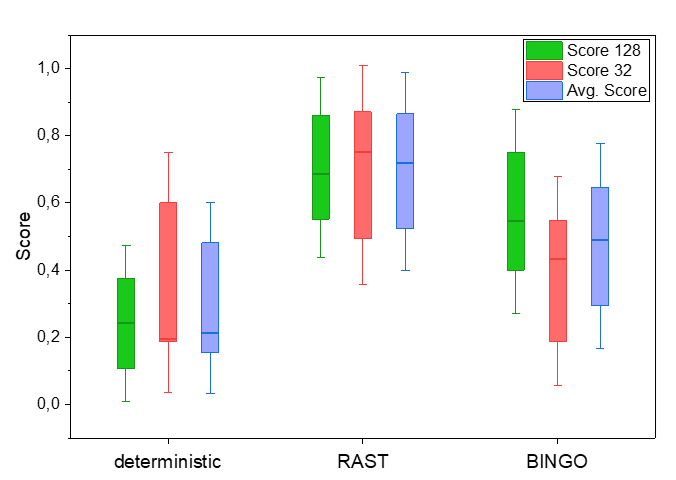}
		\captionof{figure}{%
			Metric scores for deterministic, RAST and BINGO undersampling for $128 \times 128$ and $32 \times 32$ imaging matrix and average score.}
		\label{fig:image_metrics}
	}
\end{figure}

The potential of RAST and BINGO is evident in fig.\,\ref{fig:image_metrics} through their substantially higher metric performance scores compared to deterministic undersampling. RAST achieves the highest and most consistent performance across all imaging matrix sizes, yielding improvements of 183\% for the $128\times 128$ matrix and 294\% for the $32\times 32$ matrix. This results in an average metric performance score of 0.71, corresponding to a 238\% increase over deterministic undersampling. BINGO also demonstrates substantial performance gains, with score improvements of 125\% for the larger matrix size and 104\% for the smaller matrix size, leading to an average score of 0.49 and an overall improvement of 133\% relative to deterministic undersampling across the evaluated conditions. Deterministic undersampling shows a decreasing trend in performance score with increasing undersampling factors, independent of the reconstructed matrix size. A similar trend is observed for both RAST and BINGO, as their scores also decline at higher undersampling levels. However, both introduced methods consistently outperform the deterministic undersampling, while RAST exceeds BINGO in the metric performance score by about 45\%.

\section{Discussion \& Conclusions}
Low-discrepant sampling schemes offer superior undersampling behavior compared to standard encoding schemes, such as Cartesian or radial acquisition schemes, especially in combination with CS. Moreover, data acquired with quasi-random point distributions offer additional advantages, as the reconstruction depends solely on the k-space sampling density, $\Delta k$, and can therefore be reconstructed independently of the field-of-view, FOV \cite{Speidel2}. In contrast, classical Cartesian sampling requires the k-space coverage to satisfy Nyquist’s sampling theorem, FOV $ = 1/ \Delta k$, in order to avoid aliasing artifacts in the reconstruction of the desired FOV. For quasi-random sampling schemes combined with dedicated undersampling strategies, as employed in this work, artifacts in undersampled reconstructions appear in a largely incoherent, noise-like form. This property allows reconstruction of arbitrary FOVs, theoretically limited only by noise–like artifacts, which can be effectively mitigated using techniques such as CS \cite{Speidel2}. \\ In this work, we present two algorithms for point distribution optimization for 2D SPI, which prove to yield an effective imaging approach, particularly due to the preservation of low-coherence aliasing artifacts compared to a deterministic undersampling approach. Furthermore, the proposed point distribution design can be combined with multi-coil acquisitions, where multiple receiver elements might enable additional scan-time acceleration in combination with appropriate parallel imaging or joint reconstruction techniques.

Limitations of the presented BINGO approach may include the numerical stability of the introduced criterion, as an inversion of the matrix $A$ is involved. A prerequisite of the approach, especially for the matrix determinant lemma, is the invertibility of this matrix, which might be violated for higher undersampling factors. Regularization or usage of the Moore-Penrose inverse might mitigate the issue, leading to still valuable results, however, this has been kept in mind during the optimization of point distributions. Further, choosing a particular prior $p(\vec{\rho})$ could also result in improved outcomes for a specific application, making it a good starting point for future research.

However, the advantage of the BINGO approach is its broad applicability. Since only the encoding matrix is used for the calculation of the criterion, additional features like coil sensitivities or non-linear acquisition schemes can be easily incorporated into this approach. In principle, nonlinear encoding involves arbitrary encoding fields such that resolution becomes spatially-dependent, which can also be interpreted by introducing local k-spaces instead of one global one. Hence, geometrical approaches may work well for conventional MRI, however, as soon as local k-spaces are considered, complexity can increase dramatically, making the proposed algorithm a suitable candidate for non-linear encoding strategies. Together with the potential of non-linear encoding schemes to accelerate image acquisition, further reduction of the number of encodings might be beneficial for time-critical applications.

The demonstrated acceleration of 2D SPI/CSI offers broad applicability for metabolic studies, especially in the focus of recent advances in easy-to-use hyperpolarization devices \cite{nagel2023PHIP, gierse23fumarate}. This is especially relevant for applications like Z-OMPD \cite{zompd} or thermometry \cite{thermometryFranz}, which utilize the sensitivity of the chemical shift to variations in pH or temperature. In these settings, conventional methods for spatial-spectral metabolite separation, including IDEAL \cite{ReederIDEAL, C13IDEAL} and spectral-spatial (SPSP) excitation pulses \cite{CUNNINGHAMSPSP}, are not optimal, since the resonance frequencies are not known a priori. In contrast, the presented approach acquires the full spectrum, offering the reconstruction of spectrally separated metabolite images without requiring prior knowledge of the underlying resonance frequencies.

By adding a conventional readout gradient, the proposed point distributions may be applied to rapid 3D encoding as suggested earlier for Poisson disk distributions \cite{PoissonUS} or radial phase encoding\cite{RadialUS, GRadialUS}. Here, especially for small and anisotropic matrix sizes, the proposed point distributions may be advantageous.




\section*{Acknowledgments}
This work has received funding from the European Union's Horizon 2020 research and innovation program under grant agreement No. 858149 and by German Federal Ministry of Research, Technology and Space under grants  13N16446 and 03ZU2110EA.
The authors thank the Ulm Technology Center ULMTeC for its support. The continuous support from Philips Healthcare is highly appreciated by the authors. \\

\bibliography{literature.bib}
\bibliographystyle{unsrt}

\end{document}